\documentclass[12pt]{article}
\usepackage{graphicx}
\large

\title{\large Peculiar spaces of relativistic quantum fields}

\vspace{10 cm}

\author{ V.V. Belokurov$^{1,2}$ and E.T. Shavgulidze$^{1}$    \\
\\    {\small \em 1. Lomonosov Moscow State University, Russia }
\\    {\small \em 2. Institute for Nuclear Research of Russian Academy of Sciences, Russia }
\\ {\small  vvbelokurov@yandex.ru ; shavgulidze@bk.ru}}

\date{ \ \ \  }
\begin{document}
\maketitle

\begin{abstract}

We study how quantum field theory models are modified under the reparametrizations of the space-time coordinates and some simultaneous transformations of the field function. The transformations that turn the action of the massive field in the Minkowski space-time into the action of the massless field in some curved space are presented.
\end{abstract}

\vspace{0.5cm}

\textbf{1. Introduction. }

The behaviour of functional integrals in quantum field theory under transformations of coordinates of the space-time and field functions is an interesting and important problem. In this paper, we study how the quantum field theory model is modified under the reparametrizations of the space-time coordinates $x^{\mu}$ and some simultaneous transformations of the field function $u(x)\,.$

It is known that the properties of the Wiener measure remain valid under reparametrizations of the time variable $t\,.$ In this case, there is the invariant differential $\left(u(t)\right)^{-2}dt\,.$
We consider the class of transformations of the coordinates of the $d-$ dimensional space and the simultaneous transformations of the field function $u(x)$ that leave invariant the differential
$$
 \left(u(x)\right)^{2d-4}dx\,, \ \ \ \ \ x\in \textbf{R}^{d}\,.
 $$
 In this class, there are transformations that relate massive and massless theories. If we start with the action of the massive field in the Minkowski space-time then we get the action of the massless field in some curved space. The metric tensor of this space is determined by the mass of the initial field and the form of the transformations. The metric and the curvature are singular at some points of the space, although the determinant of the metric does not change: $\det G_{\mu\nu}=-1\,.$

\vspace{0.5cm}

\vspace{0.5cm}

\textbf{ 2. Diffeomorphisms of the time variable and generation of the masses of quantum particles.}

\vspace{0.5cm}

In \cite{(BSh)} we have
 presented the explicit form of the diffeomorphisms of the time variable
 $\,t\rightarrow g(t)\,;\ \ g'(t)>0\,$
 that transform the measure for quantum free massless particle into the measure for quantum free massive particle .

We have used the quasiinvariance of the Wiener measure
on the space of continuous functions
under the group of diffeomorphisms
$$
\int\limits_{C\left([a,\,b]\right)} F[y]\,\exp \left\{-\frac{1}{2}\int \limits _{a}^{b}\left(y'(\tau)\right)^{2}\,d\tau\right\}\,dy=
$$
\begin{equation}
   \label{1.1}
   C(g)\,
\int\limits_{C\left([a,\,b]\right)} F[gx]\,
\exp \left\{\frac{1}{4}\int \limits _{a}^{b}\,x^{2}(t)\mathcal{S}_{g}(t)\,dt+\frac{1}{4}\left(x^{2}(b)\frac{g''(b)}{g'(b)} -x^{2}(a)\frac{g''(a)}{g'(a)}\right)\right\}\,
\end{equation}
$$
\exp \left\{-\frac{1}{2}\int \limits _{a}^{b}\left(x'(t)\right)^{2}\,dt\right\}\,dx\,.
$$

The function  $y(\tau)$ is the image of the function $x(t)$  realizing the following representation of the group of diffeomorphisms
\begin{equation}
   \label{1.2}
y(\tau)=(gx)(\tau)=x(t)\sqrt{g'(t)}=x\,\left(g^{-1}(\tau)\right)
\,\frac{1}{\sqrt{\left(g^{-1}\right)'(\tau)}}\,.
\end{equation}

Here $\mathcal{S}_{g}$ is the Schwarz derivative
$$
\mathcal{S}_{g}(t)\equiv
\left(\frac{g''(t)}{g'(t)}\right)'
-\frac{1}{2}\left(\frac{g''(t)}{g'(t)}\right)^2\,.
$$

 For the diffeomorphisms of the form
\begin{equation}
   \label{1.3}
g_{0}(t)=\frac{1}{\left(e^{2mb}-e^{2ma}\right)}\left\{(b-a)\,e^{2m t}+(a\,e^{2mb} - b \,e^{2ma} ) \right\}
\end{equation}
and
\begin{equation}
   \label{1.4}
g(t)=\frac{\left(a+b+\delta \right)g_{0}(t)-ab}{g_{0}(t)+\delta}
\end{equation}
 where $\delta$ is an arbitrary parameter,
 the Schwarz derivative is the constant
$$
\mathcal{S}_{g}(t)=-2\,m^{2}\,.
$$

The corresponding coefficient is
\begin{equation}
   \label{1.5}
C(g)=\frac{\sinh \left(m(b-a)\right)}{m(b-a)}\,.
\end{equation}
For convenience, we suppose that  $x(a)=x(b)=y(a)=y(b)=0\,.$

Thus, we have
$$
\int\limits_{C\left([a,\,b]\right)} F[y]\,\exp \left\{-\frac{1}{2}\int \limits _{a}^{b}\left(y'(t)\right)^{2}\,dt\right\}\,dy=
$$
\begin{equation}
   \label{M7}
C(g)\,\int\limits_{C\left([a,\,b]\right)} F[gx]\,
\exp \left\{-\frac{1}{2}\int \limits _{a}^{b}\,m^{2}x^{2}(t)\,dt\ -\frac{1}{2}\int \limits _{a}^{b}\left(x'(t)\right)^{2}\,dt\right\}\,dx\,.
\end{equation}

This equation means that \textbf{quantum theory of a free massive particle is nothing more than quantum theory of a free massless particle but with the different way of the measurement of time and the corresponding change of canonical coordinates} $x\,.$

The direct realization of the above scheme in local quantum field theory faces a serious obstacle.
  In fact, the three-dimensional Fourier transform of the free field function represents the infinite set of quantum harmonic oscillators with the frequencies   $\sqrt{\vec{k}^{2}} $ and  $\sqrt{\vec{k}^{2}+m^{2}}\,, $ for the massless and the massive fields, respectively.

Having in mind the infinite time interval we consider the diffeomorphism
\begin{equation}
   \label{1.11}
g_{{}_{m}}(t)=\exp\{2mt\}\,.
\end{equation}
Its Schwarz derivative equals to $-2m^{2}\,,$ and the inverse diffeomorphism has the form
\begin{equation}
   \label{1.12}
g_{{}_{m}}^{-1}(t)=\frac{1}{2m}\ln t\,.
\end{equation}
In this case, the transfer from a massless relativistic field to the massive relativistic field is realized by the composition of the diffeomorphisms of the time coordinate
\begin{equation}
   \label{1.13}
g_{{}_{\sqrt{\vec{k}^{2}+m^{2}}}}\circ g_{{}_{\sqrt{\vec{k}^{2}}}}^{-1}(t)\,.
\end{equation}
The explicit form of the composition  (\ref{1.13}) is
\begin{equation}
   \label{1.14}
\tau _{k}=g_{{}_{\sqrt{\vec{k}^{2}+m^{2}}}}\circ g_{{}_{\sqrt{\vec{k}^{2}}}}^{-1}=g_{{}_{\sqrt{\vec{k}^{2}+m^{2}}}}\left( g_{{}_{\sqrt{\vec{k}^{2}}}}^{-1}(t)\right)=t^{\sigma (k)}\,,\ \ \ \sigma (k)=\frac{\sqrt{\vec{k}^{2}+m^{2}}}{\sqrt{\vec{k}^{2}}}\,.
\end{equation}

Thus, for free scalar field we have
$$
\exp\{-\frac{1}{2}\int d^{3}\vec{k}\,\int dt\,\left(|\dot{\phi}(t,\vec{k})|^{2}+\vec{k}^{2}\,|\phi(t,\vec{k})|^{2} \right)\}\,d\phi=
$$
\begin{equation}
   \label{1.15}
\exp\{-\frac{1}{2}\int d^{3}\vec{k}\ Z(\vec{k})\,\int d\tau_{k}\,\left(|\dot{\varphi}(\tau_{k},\vec{k})|^{2}+\left(\vec{k}^{2}+m^{2}\right)|\varphi(\tau_{k},\vec{k})|^{2} \right)\}\,d\varphi\,.
\end{equation}

As the factor $Z(\vec{k})$ depends on $\vec{k}$, the right-hand side of the equation
 (\ref{1.15}) corresponds to a nonlocal theory. In the next sections we develop another approach that makes it possible to transform the massive field theory to the massless one.

\vspace{0.5cm}

\textbf{3. Mass of relativistic field and deformation of the geometry of space-time.}

\vspace{0.5cm}

Consider the action of self-interacting scalar field in the ordinary Minkowski\footnote {The similar scheme can also be carried out in the Euclidean space.} space-time
\begin{equation}
   \label{3.1}
   2\mathcal{A}=\int \left [\left(\frac{\partial u}{\partial x^{0}} \right)^{2}- \left(\frac{\partial u}{\partial x^{i}} \right)^{2} - m^{2}u^{2} + u^{4}\right]\,d^{4}x\,.
\end{equation}
Representing the field $u(x)$ as the product
$$
u(x)=v(x)\,\varphi (x)
$$
and supposing that $u$ and $v$ vanish at infinity, we rewrite
the action (\ref{3.1}) in the form
$$
 2\mathcal{A}=\int \left [\left(\frac{\partial v}{\partial x^{0}} \right)^{2}- \left(\frac{\partial v}{\partial x^{i}} \right)^{2}\right]\,\varphi^{2}\,d^{4}x\,
$$
\begin{equation}
   \label{3.2}
     +\, \int \, v^{4}\,\varphi^{4}\,d^{4}x\,
   +\,\int \,\varphi\, \left [-\frac{\partial ^{2}\varphi}{(\partial x^{0}) ^{2}}+ \frac{\partial ^{2}\varphi}{(\partial x^{i})^{2}} - m^{2}\varphi \right]\,v^{2}\,d^{4}x\,.
   \end{equation}

If the function $\varphi (x)$ satisfies the KG equation
\begin{equation}
   \label{3.3}
     -\frac{\partial ^{2}\varphi}{(\partial x^{0}) ^{2}}+ \frac{\partial ^{2}\varphi}{(\partial x^{i})^{2}} - m^{2}\varphi =0\,,
   \end{equation}
then the action (\ref{3.2}) is reduced to the action of the massless field in some curved space
\begin{equation}
   \label{3.4}
   \int \left [G^{\mu\nu} \frac{\partial w}{\partial \xi ^{\mu}} \frac{\partial w}{\partial \xi ^{\nu}}+w^{4}\right] \sqrt{-G}\,d^{4}\xi\,,
\end{equation}
where $v(x)=w(\xi (x))\,.$ The geometry of the space is determined by the solution of the KG equation $\varphi (x)\,.$

 We demand the form of the interaction term to be invariant. In this case, the Jacobian of the substitution
 \begin{equation}
   \label{3.5}
   \det \left(\frac{\partial \xi}{\partial x} \right) =\varphi^{4}(x)\,.
  \end{equation}
   and
   $$
   G=\det G_{\mu\nu}=-1\,.
   $$
   However the metric tensor $G^{\mu\nu}$ is nontrivial.

There are various options for the function $\xi^{\mu} (x^{\nu})\,.$ First, let us consider a special case with $\varphi=\varphi (s)\,,\ $ where $s=\sqrt{(x^{0})^{2}-(x^{i})^{2}}\,.$ The KG equation is transformed to the ordinary differential equation
\begin{equation}
   \label{3.6}
    \varphi ''+\frac{3}{s}\varphi '+ m^{2}\varphi =0\,.
\end{equation}
Its general solution is expressed in terms of the Bessel functions
\begin{equation}
   \label{3.7}
    \varphi (s)=\frac{1}{s}\left[C_{1}J_{1}(ms)+C_{2}Y_{1}(ms) \right]\,.
   \end{equation}
The values of the constants $C_{1},\ C_{2}$ are determined by the boundary conditions. In particular, $ s^{-1} J_{1}(ms)\sim const$ and
$s^{-1} Y_{1}(ms)\sim C\left(\ln s-2m^{-2}s^{-2}\right)$ at $s\sim 0\,.$

A simple but nontrivial way to define the new coordinates is a dilatation of the old ones with the dilatation factor depending on $s$
$$
\xi^{\mu}= \rho(s)s^{-1}\,x^{\mu}\,,\ \ \ \rho(0)=0\,.
$$
In this case, the angles do not change and
$$
\sigma \equiv \sqrt{(\xi^{0})^{2}-(\xi^{i})^{2}} =\rho(s) \,.
$$
Evaluating the Jacobian of the substitution and taking into account (\ref{3.5}) we get the  equation for $\rho (s)$
\begin{equation}
   \label{3.8}
   \rho ^{3}(s) \rho'(s)s^{-3} =\varphi ^{4}(s)\,.
   \end{equation}
   Thus,
   $$
   \rho^{4}(s)=\int \limits _{0}^{s^{4}}\,\varphi ^{4}(s) ds^{4}\,.
   $$

The points of the $x$ space where $\varphi (s)=0$ correspond to some singular points in $\xi$ space. To explain this statement in more detail, in the next section we consider a slightly different substitution $\xi^{\mu} (x^{\nu})\,.$

\vspace{0.5cm}

\textbf{4. The structure of the peculiar spaces of relativistic quantum fields for some special diffeomorphisms of space-time coordinates.}

\vspace{0.5cm}

Now, we suppose that the function $\varphi$ does not depend on the spatial variables:  $\varphi=\varphi(t)\,.$
In this case, the KG equation reduces to the equation for a harmonic oscillator with the solution
\begin{equation}
   \label{4.1}
   \varphi(t)=\mu \sin m(t-t_{0})\,.
   \end{equation}

Consider the following transformations of space-time
\begin{equation}
   \label{4.2}
 \tau=f(t)\,,\ \ \ \xi^{i}=f'(t)\,x^{i} \,,
   \end{equation}
and let $ \varphi(t)$ be equal to $f'(t)$
\begin{equation}
   \label{4.3}
 u(t,x)=v(t,x)\,\varphi(t)=w\left(f(t),\,f'(t)\,x\right)\,f'(t)\, .
   \end{equation}

In order that the directions of the initial time $t$ and the new time $\tau$ be the same, the function $\varphi(t)= f'(t)$ must be nonnegative.
However, the solution (\ref{4.1}) does not satisfy this condition. To overcome the problem, note that at the points $t^{\ast}$ where $\varphi(t)=0$
it may not be a solution of the equation (\ref{3.3}). If the zeroes of the function $\varphi(t)$ form a discrete set then they do not give any contribution to the integral. So, we can take arbitrary solutions on the intervals between the  zeroes and sew them together at these points.

For simplicity, we take
\begin{equation}
   \label{4.4}
   \varphi(t)=\mu |\sin m(t-t_{0})|\,.
   \end{equation}

We denote by $g(\tau)$ the diffeomorphism inverse to $f(t)$
$$
t=g(\tau)=f^{-1}\left(f(t)\right)\,,
$$
and recall some useful relations
$$
f'(t)=\frac{1}{g'(\tau)}\,, \ \ \ \frac{f''(t)}{f'(t)}=-\frac{g''(\tau)}{\left(g'(\tau)\right)^{2}}\,.
$$
The action (\ref{3.1}) is written in the form
$$
 \int \left [G^{\mu\nu} \frac{\partial w}{\partial \xi ^{\mu}} \frac{\partial w}{\partial \xi ^{\nu}}+w^{4}\right] \sqrt{-G}\,d^{4}\xi
$$
   \begin{equation}
   \label{4.5}
 =\int \left [\left( \frac{\partial w}{\partial \tau}-h(\tau)\,\xi^{i}\frac{\partial w}{\partial \xi ^{i}}\right)^{2}-\left( \frac{\partial w}{\partial \xi ^{i}}\right)^{2}+w^{4}\right] \,d\tau d^{3}\xi
    \end{equation}
where
$$
h(\tau)=\frac{g''(\tau)}{g'(\tau)}\,.
$$

Thus, the metric tensor is
\begin{equation}
   \label{4.6}
   G^{00}=1\,,\ \ G^{0i}=-h(\tau)\,\xi^{i}\,,\ \ G^{ij}=h^{2}(\tau)\,\xi^{i}\xi^{j} -\delta^{ij}\,.
   \end{equation}

By the direct evaluation\footnote{We are grateful to A.S. Ivanov for the help in analytic computations.}, one can check that $\det G_{\mu\nu}=-1 $ and get the expression for the Riemann tensor and the scalar curvature
\begin{equation}
   \label{4.7}
 R=2\left[3h'( \tau)+ \left(\xi_{1}^{2}+\xi_{2}^{2}+\xi_{3}^{2}\right)\left(h'(\tau) \right)^{2}+ \left(\xi_{1}^{2}+\xi_{2}^{2}+\xi_{3}^{2}\right)h(\tau)h''(\tau) \right]\,.
 \end{equation}

Now, let us find the explicit form of the dependence $\tau=f(t)\,,\ t=g(\tau)\,,\ h(\tau)\,.$
From (\ref{4.4})
\begin{equation}
   \label{4.8}
\tau=f(t)=\frac{\mu}{m}\int\limits_{0}^{t} |\sin mt|\,m\,dt\,.
 \end{equation}
Here, we put the constant $t_{0}=0$ and assume that $\tau=0$ at $t=0\,.$

We integrate (\ref{4.8}) over intervals $(k-1)\frac{\pi}{m}\leq t \leq k\frac{\pi}{m}$ separately and sew the results together.

At the interval $(k-1)\frac{\pi}{m}\leq t \leq k\frac{\pi}{m}\,, \ \ \ k=1,2,...\,,$  the result is
\begin{equation}
   \label{4.9}
\tau=\frac{\mu}{m}\left((-1)^{k}\cos mt+2k-1 \right)\,, \ \ \ 2(k-1)\frac{\mu}{m}\leq  \tau \leq 2k\frac{\mu}{m}\,.
 \end{equation}
In this case,
\begin{equation}
   \label{4.10}
t=g(\tau)=-\frac{1}{m}\arccos \left(\tau-(2k-1)\frac{\mu}{m} \right)+k\frac{\pi}{m}\,,
 \end{equation}
 \begin{equation}
   \label{4.11}
\frac{1}{g'(\tau)}=m\,\sqrt{\left(\tau-(2k-1)\frac{\mu}{m} \right)\,\left(-\tau+2k\frac{\mu}{m} \right)}\,,
 \end{equation}
and
\begin{equation}
   \label{4.12}
2h(\tau)=\frac{1}{\tau-(2k-1)\frac{\mu}{m} }-\frac{1}{2k\frac{\mu}{m}-\tau} \,.
 \end{equation}

Thus, the function $\tau=f(t)$ is a continuous monotone increasing one with the points of inflection at $t= k\frac{\pi}{2m}\,.$
Note that it is valid for negative values of $t$ and $\tau$ as well.

At some values of the new time variable, namely at $\tau=k\frac{\mu}{m}\,,$ the components of the metric tensor $G^{0i},\, G^{ij}$ and the scalar curvature $R$ tend to infinity.
On the other hand, $\frac{1}{g'(\tau)}\,,$ and hence the space coordinates $\xi^{i}\,,$ vanish at these singular points. It makes sense to emphasize that the points  $\tau=k\frac{\mu}{m}\,,$ where the space collapses, in this case are determined only by the mass of the field $m\,.$  

One can easily check that the form of the action of a gauge field is invariant at the transfer to the new space. In fact, in the Minkowski space the action of the electromagnetic field is
\begin{equation}
   \label{4.13}
\mathcal{A}_{em}=-\frac{1}{4}\int\,\eta^{\alpha\gamma}\eta^{\beta\delta}f_{\alpha\gamma}f_{ \beta\delta}\,d^{4}x\,, \ \ \ \ \ f_{\alpha\gamma}=\frac{\partial a_{\alpha}}{\partial x^{\gamma}}-\frac{\partial a_{\gamma}}{\partial x^{\alpha}}\,.
 \end{equation}
Using the following relations
\begin{equation}
   \label{4.14}
\frac{\partial }{\partial x^{\gamma}}=\frac{1}{g'}V^{\lambda}_{\gamma}\frac{\partial }{\partial \xi^{\lambda}}\,, \ \ \ \ \ a_{\alpha}=\frac{1}{g'}V^{\lambda}_{\alpha}A_{\lambda }\,,\ \ \ \ \ \ V^{\lambda}_{\gamma}=g'\,\frac{\partial \xi^{\lambda} }{\partial x^{\gamma}}\,.
 \end{equation}
one gets
 \begin{equation}
   \label{4.15}
\mathcal{A}_{em}=-\frac{1}{4}\int\,G^{\mu\lambda}G^{\nu\sigma}F_{\mu\lambda}F_{ \nu\sigma}\,\sqrt{-G}\,d^{4}\xi\,, \ \ \ \ \ F_{\mu\lambda}=\frac{\partial A_{\mu}}{\partial \xi^{\lambda}}-\frac{\partial A_{\lambda}}{\partial \xi^{\mu}}\,,
 \end{equation}
were
 \begin{equation}
   \label{4.16}
   G^{\mu\lambda}=\eta^{\alpha\gamma}\,V^{\mu}_{\alpha}\,V^{\lambda}_{\gamma}\,.
 \end{equation}

The same result is valid for nonabelian gauge fields as well.

\vspace{0.5cm}

\textbf{5. Discussion.}

Now we do not know if the points, where the discussed peculiar spaces collapse, have any physical sense.
However, we hope that the study of the different spaces, where quantum fields live, can help to understand the structure of functional integrals in quantum field theory. 

The point is that
functional integrals do not change at the transformations. Moreover, as the field function substitution is linear, the Gaussian integrals transfer to  the Gaussian integrals.
In the case considered, the factor in the field function transformation (\ref{4.3}) is singular at some points. Therefore, the functional integrations in the theories  in the initial and the new spaces are carried out over different functional space. A simple example of the modification of the functional space at nonlinear nonlocal substitutions in functional integrals was given in \cite{(BSh2)} in some other context.

\end{document}